\documentclass[iop,apj]{emulateapj}
\usepackage{natbib}

\begin{document}

\title{The EUV emission from sun-grazing comets}       
\author{P. Bryans$^1$ and W. D. Pesnell$^2$}
\affiliation{$^1$ ADNET Systems Inc., NASA Goddard Space Flight Center, Code 671, Greenbelt MD 20771, USA}
\affiliation{$^2$ NASA Goddard Space Flight Center, Code 671, Greenbelt MD 20771, USA}



\begin{abstract}

The Atmospheric Imaging Assembly (AIA) on the Solar Dynamics Observatory (SDO) 
has observed two sun-grazing comets as they passed through the solar atmosphere.
Both passages resulted in a measurable enhancement of Extreme Ultraviolet
(EUV) radiance in several of the AIA bandpasses. We explain this EUV emission by
considering the evolution of the cometary atmosphere as it interacts with the
ambient solar atmosphere. 
Molecules in the comet
rapidly sublimate as it approaches the Sun. They are then photodissociated
by the solar radiation field to create atomic species.
Subsequent ionization of these atoms produces a higher abundance
of ions than normally present in the corona and results in EUV
emission in the wavelength ranges of the AIA telescope passbands.

\end{abstract}


\section{Introduction}
\label{sec:intro}

Thousands of sun-grazing comets have been detected since the advent of
space-based solar observation.  The most successful instrument for observing these
has been the Large Angle and Spectrometric Coronagraph \citep[LASCO;][]{lasco} on the
Solar and Heliospheric Observatory \citep[SOHO;][]{soho}, which has observed
more than 2000 such comets approaching the Sun 
\citep[see, e.g.,][]{biesecker02, knight10}.  For a complete catalogue of sun-grazing comets, see the US
Naval Research Laboratory {\em Sungrazing Comets}
website.\footnote{http://sungrazer.nrl.navy.mil/} Very few sun-grazing
comets have been seen to survive their close passage to the solar photosphere and
emerge post perihelion \citep{marsden05}, none of which have been observed by LASCO
until 2011.\footnotemark[1]

Despite the lack of direct observation of the destruction of a sun-grazing comet,
the literature is not lacking in predictions of the results of such an event.
It is not possible to give a full literature review here, but we highlight some papers of
particular relevance.
\citet{weissman83} and \citet{sekanina84} discussed the physical processes of import to the
destruction of comets during their perihelion passage.  A more complete model detailing the
erosion of sun-grazing comets was presented by \citet{sekanina03}.
And, most recently, \citet{brown11} have provided analytical models describing the
destruction mechanism of sun-grazing and -impacting comets. This latter work indicates that
the dominant mass-loss mechanism varies between sublimation, ablation, and explosion
depending on the cometary mass and perihelion distance.
One particularly interesting prediction is that a sun-impacting comet large enough to reach
the chromosphere would result in solar flare-like energy release.

Further from the Sun, it has been known for around 15 years that comets emit at X-ray and EUV
wavelengths. In 1996, comet C/1996 B2 (Hyakutake) was observed
by the ROSAT X-ray telescope \citep{lisse96}. Subsequently, all comets within 3~AU
were found to emit X-rays \citep{dennerl97}.  The discovery of this emission was
initially surprising, given the low temperature of the cometary atmosphere, but was
explained as arising from charge exchange between solar
wind ions and neutral cometary species \citep{cravens97, krasnopolsky97}. However,
until 2011, there had been no observations of a sun-grazing comet close to
perihelion, let alone at such wavelengths.

The launch of the Solar Dynamics Observatory \citep[SDO;][]{sdo} in 2010 provides a
unique capability to observe sun-grazing comets. The
Atmospheric Imaging Assembly \citep[AIA;][]{aia} 
images the entire
Earth-facing solar corona at a cadence of 12~s with  $\sim 1.2$~arcsec resolution in
7 EUV channels, with passbands centered at 94, 131, 171, 193, 211, 304, and 335~\AA.
Should a sun-grazing comet be sufficiently bright in the EUV
as it flies within the field-of-view of AIA,
the high spatial and temporal resolution of AIA makes it the 
ideal instrument to observe such an event.

The first detection of a comet by SDO was on 2011 July 5-6, when comet C/2011 N3
(SOHO) passed across the disk of the Sun \citep{schrijver12}. It was detected in
several of the AIA passbands before disappearing as it evaporated in the
solar atmosphere.  A second comet, C/2011 W3 (Lovejoy), was observed on 2011
December 15-16, this time passing behind the east limb of the solar disk and emerging on the west
limb.  Again, several of the AIA passbands detected EUV emission. As we will show below,
cometary neutral species cannot survive very long at such close proximity to the solar radiation
field. Thus, the explanation for EUV emission in sun-grazing comets must differ from the
charge-exchange model used to explain the X-ray emission detected in the
heliosphere. This paper offers such an explanation.

The remainder of the paper is organized as follows.
In Section~\ref{sec:observations} we describe the AIA observations of the two
comets.
In Section~\ref{sec:model} we describe the model of the comet-corona interaction
used to explain the EUV emission, including our simplified description of the
cometary topology and the physical parameters of the cometary and solar
atmospheres used in our calculations. Section~\ref{sec:material} outlines the
rate of evolution of the cometary material as it is sublimated from the comet
body and subsequently dissociated and ionized in the corona. In
Section~\ref{sec:results} we show what contribution the cometary material has to
EUV emission in the AIA channels.  
Finally, in Section~\ref{sec:discussion} we give our conclusions and
offer some suggestions on future work to improve our
emission model.

\section{Observations}
\label{sec:observations}

\subsection{Comet C/2011 N3 (SOHO)}

Comet C/2011 N3 (SOHO) was first detected by the LASCO coronagraph in white light,
approaching the west limb of the Sun. On 2011 July 5-6 it passed across the disk of
the Sun and was detected by AIA \citep{schrijver12}. 
The first detection was in AIA 171~\AA, off the solar limb, at 2011 July 5 23:46 UT,
and was tracked until 2011 July 6 00:05:50 UT. Emission was also detected in the 131,
193, 211, and 335~\AA\ passbands. The observations suggest that the comet may
have fragmented before evaporating in the solar atmosphere.

Of the 7 EUV channels of AIA, 5 show a response to the comet \citep{schrijver12}.
There was no significant detection in the 94 or 304~\AA\ channels; the 94~\AA\
channel has a low signal-to-noise ratio in these observations and the 304~\AA\
channel is dominated by the bright emission from lower in the solar
atmosphere. The enhancement over the background emission is greatest in the 131 and
171~\AA\ passbands, where the cometary emission is $\sim 10$\% above the background.  However, the comet
brightness was determined by summing over a box of $30\times 15$ AIA pixels---a
significantly larger area than the emitting region---so the true intensity
enhancement due to the comet is greater.

\subsection{Comet C/2011 W3 (Lovejoy)}

A second comet, C/2011 W3 (Lovejoy), was observed by SDO on 2011 December 15. This
passage of another Kreutz-family comet 5 months later resulted in perihelion being on the far side of the Sun
from the perspective of SDO. Ground observations several days prior to perihelion
gave an accurate estimate of the cometary orbit \citep{green11}, allowing SDO to repoint 1250~arcsec
east of solar center, giving extended observations of the
comet as it approached the eastern limb. Comet C/2011 W3 (Lovejoy) was appreciably
more massive than C/2011 N3 (SOHO), allowing it to survive perihelion and emerge on the
other side of the Sun (see
\citet{gundlach12} for a discussion on the survival).
SDO had repointed to Sun center in time to observe the
emergence.
An in-depth analysis of the orbit can be found in \citet{sekanina12}. We make particular
note of the absence of observable cometary dust at the time of the AIA observations
as a result of any dust particles being rapidly sublimated \citep{sekanina12}.

Despite perihelion passage 
being obscured by the solar disk, C/2011 W3 (Lovejoy) resulted in, perhaps,
more interesting observations than C/2011 N3 (SOHO). The
absence of contamination from background emission from the solar disk in the off-limb
observations gives a clear picture of sublimated cometary material following the
solar magnetic field.
In contrast to C/2011 N3 (SOHO), C/2011 W3 (Lovejoy) produced
significant signals in the 304~\AA\ channel. 
For the channels that also detected C/2011 N3 (SOHO), the brightness contrast is similar in the
C/2011 W3 (Lovejoy) observations.

\begin{figure*}
  \includegraphics{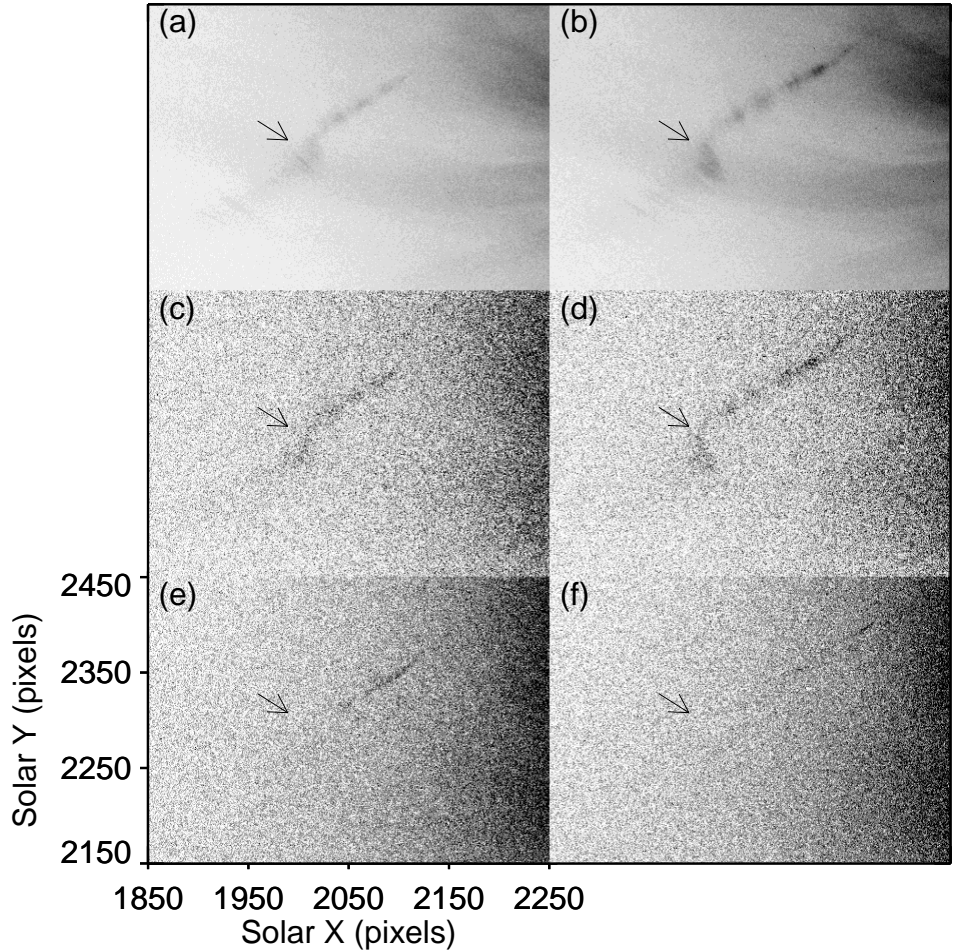}
  \caption{Emission from comet C/2011 W3 (Lovejoy) pre-perihelion.  
  The color scale has been reversed, with darker colors indicating brighter emission.
  The comet is traveling up and to the right, towards the solar disk, which is to the 
  extreme right of the images.  
  All panels show the same location.  Axes units are given in AIA pixels with (0,0)
  being the bottom left of the full $4096\times 4096$ AIA image.
  Panels (a) and (b) show emission 
  from the 171~\AA\ channel of AIA at 00:00:12~UT and 00:01:11~UT, respectively;  
  panels (c) and (d) show the 131~\AA\ channel at 00:00:10~UT and 00:01:10~UT, 
  respectively; and panels (e) and (f) show the 304~\AA\ channel 
  at 00:00:08~UT and 00:01:08~UT, respectively.
  The arrow is at the same location in each image and indicates the kink in
  the tail described in the text.}
  \label{fig:lovejoy ingress}
\end{figure*}

Fig.~\ref{fig:lovejoy ingress} shows the comet ingress in the 171, 131, and 304~\AA\
wavelength channels of AIA.  Two images at each wavelength are shown, separated by
1~minute, and nearly simultaneous across the three wavelength channels. Panels (a) and (b)
show the 171~\AA\ emission, where evidence of the cometary material following the magnetic
field is most clearly visible.  The material is seen to form clear striations, roughly
perpendicular to the direction of the comet's motion.  These striations are long-lived; the
brightest was first detected in the 171~\AA\ channel at 2011 Dec 15 23:57:47~UT and seen to
persist until 2011 Dec 16 00:16:23~UT, almost 19 minutes, and 8 minutes after the comet went
behind the solar limb.

Panels (c) and (d) of fig.~\ref{fig:lovejoy ingress} show the 131~\AA\ observations at close
to the same time as those of (a) and (b).  The signal-to-noise is smaller than in the
171~\AA\ passband, but the cometary emission is spatially similar. This is not the case for
304~\AA\ emission, shown in panels (e) and (f).  The emission shown in panel (e) is more
spatially confined than that of (a) and (c).  Comparing the emission at the later observation
time shown in panel (f) with (b) and (d) makes this even more apparent.   For ease of
comparison across the images, we have highlighted with an arrow a point where the emission
from the comet tail appears to kink due to the magnetic field of the corona.  Sunward of this
point, cometary material appears to move up and left as it follows the magnetic field.  In
contrast, the emitting material below the arrow moves down and right. We also note that there
is no detectable signal in the 304~\AA\ channel at this point at the time of these images.

The emission in the
171 and 131~\AA\  channels is far longer-lived than that of 304~\AA. Under normal Quiet Sun (QS)
conditions, these 3 channels image plasma at $\log T$ (K) of 5.9, 5.6, and 4.7 respectively.
The similarity in the emission from the 131 and 171~\AA\ channels thus indicate a different
ionization/excitation mechanism from the QS. We suggest a solution to this problem in
Section~\ref{sec:results}.

The differences in the 
comets' detections clearly indicate that characteristics of the comet
and viewing geometry
significantly influence the intensity of EUV
emission caused by their interaction with the solar atmosphere.  It is our aim in
this work to propose a mechanism whereby emission is possible; characterizing the
precise nature of the emission depends on an accurate knowledge of the cometary
and coronal parameters.

\section{Emission model}
\label{sec:model}

As the comet approaches the Sun, material is sublimated from the comet surface.
The mass loss rate and total mass lost of C/2011 N3 (SOHO)
during the visibility in the AIA images is
estimated to be (0.01 -- 1)$\times 10^8$~g/s and (0.06 -- 6)$\times 10^{10}$~g,
respectively, by \citet{schrijver12}.

\begin{figure*}
  \centering
  \vspace{-100pt}
  \includegraphics[width=6in]{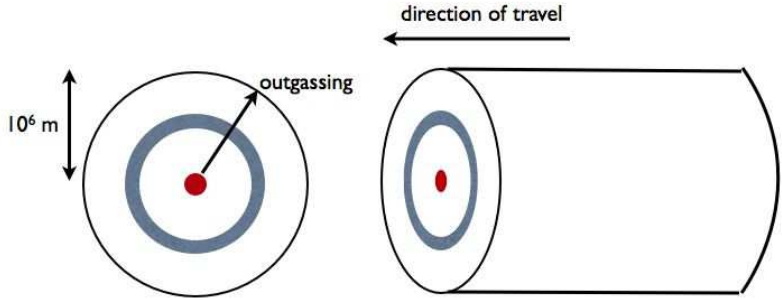}
  \vspace{-100pt}
  \caption{Schematic of the cometary emission model.  The figure on the left shows the cross
  section perpendicular to the direction of motion.  The figure on the right shows the comet
  trajectory as seen by AIA.  The red dot in the center indicates the comet body.  The blue
  area indicates one of the shells described in the text where a given ion exists.}
  \label{fig:cartoon}
\end{figure*}

We have constructed a simplified model that describes the evolution of the
cometary material as it interacts with the solar atmosphere.  We represent the
comet coma as a cylinder along the direction of travel.  A schematic
representation is shown in Fig.~\ref{fig:cartoon}. The body of the comet is
represented by the red circle at the center with radius $r_c$.  
We use $r_c=50$~m, the upper limit of the nucleus radius given by \citet{schrijver12}.
The material
sublimated from the surface then expands into the corona and forms the
cylindrical geometry shown, with axis along the path of the comet's motion. The
atoms that are formed in the coma become more highly ionized as they expand into
the corona.  We represent the space occupied by these ions as concentric shells
(shown in blue in Figure~\ref{fig:cartoon}) with inner radius $r_1$ and outer
radius $r_2$.  These radii are determined by the dissociation and ionization
rates and the outgassing velocity.

The biggest failing of this model is that we do not take account of the magnetic field.
Rather than a simple radial expansion, once the cometary material is ionized, it will be
affected by the coronal magnetic field.  As the level of ionization increases, the Larmor
radius of the ions will decrease and the more highly charged species will form tighter
spirals around the field lines.  This is somewhat in opposition to the model presented here.
The observations of C/2011 W3 (Lovejoy) clearly show that the cometary material is
influenced by the coronal magnetic field, but whether the passage of the comet has any
significant bearing on the underlying field has yet to be determined.  Accounting for the
magnetic field is not trivial and we leave such a study to a future work.
However, the ions will move in a cylindrical pattern.

The precise composition of the material sublimated from the comet's surface is
unknown.  For the purposes of this work, we adopt the elemental abundances of
Comet 1P/Halley (given in Table~\ref{table:abund}) from \citet{delsemme88}. H and O
are the most abundant elements due largely to the presence of water ice in the
comet and oxides being dominant in the asteroidal dust particles.  
Once it is deposited within the corona, H is rapidly ionized and thus does not contribute in any significant
degree to the EUV emission.   The most important elements to consider are O and
Fe; O has a high abundance in the comet and 
many of the AIA channels cover wavelength ranges with strong Fe emission lines.
To be as complete as possible, we have examined the 8 most abundant elements of
Comet 1P/Halley, i.e., those listed in Table~\ref{table:abund}---H, C, N, O, Mg,
Si, S, and Fe.  We will show that, of these, only O and Fe cause any significant
emission in the AIA passbands.

\begin{deluxetable}{cc}
\tablecolumns{2}
\tablewidth{0pt}
\tablecaption{Elemental Abundance of Comet}
\tablehead{
\colhead{Element} &
\colhead{Fractional Abundance (Number)}}
\startdata
H  & 0.484 \\
C  & 0.137 \\
N  & 0.023 \\
O  & 0.304 \\
Mg & 0.011 \\
Si & 0.016 \\
S  & 0.010 \\
Fe & 0.011
\enddata
\tablecomments{Elemental abundances in Comet 1P/Halley \citep{delsemme88}.}
\label{table:abund}
\end{deluxetable}

There is some uncertainty in the outflow velocity from the surface; from the observations of
C/2011 N3 (SOHO), we estimate the time taken for material to  travel from the surface of the
nucleus to the extent of the visible coma to be $\sim60$~s.  Assuming the material flows
radially from the nucleus at a constant velocity gives an outflow velocity of $v_{\rm
out}=17$~km~s$^{-1}$.  We explore the sensitivity of the model to this time scale
by allowing it to vary from 10~s to 90~s. The emission predicted by the model described here is
dependent on the total time the cometary material remains emitting, rather than on the
precise outflow velocity, so any influence of the magnetic field on the direction and
magnitude of $v_{\rm out}$ will not greatly affect the model results if the emitting time
remains unchanged.  For comet C/2011 W3 (Lovejoy), however, emission is seen to persist for
up to 20 minutes in certain AIA channels.  The nature of this emission is somewhat different
from the simplistic model presented here. It is seen to follow the solar magnetic field but,
after the first minute of emission, does not appear to expand into the corona, rather
maintaining consistent spatial dimensions before gradually fading from view.  For this
reason, we have chosen to use the outflow velocities of $v_{\rm out}=17$~km~s$^{-1}$ for the
emission calculations, but we make some comments on the extended duration of the C/2011 W3
(Lovejoy) emission in Section~\ref{sec:results}.

The comet perihelion parameter of $q=0.0052986$ \citep{schrijver12}
for C/2011 N3 (SOHO) corresponds
to a height above the solar surface of 97,200~km. 
Both comets were observed to pass through regions of low solar activity.
Typical values for the QS electron temperature and density at this height are $T_{\rm e}=1.5\times
10^6$~K and $10^8$~cm$^{-3}$, respectively.  We use these
values throughout for the zone of interaction between the comet and corona. In
the absence of the comet, emission measured by AIA comes predominantly from lower
in the atmosphere.  When comparing observations of the comet with background
QS emission, we use an average QS differential emission measure
(DEM) from \citet{vernazza78} and an electron density of $5\times
10^8$~cm$^{-3}$.

\section{Dissociation and ionization of cometary material}
\label{sec:material}

\subsection{Dissociation}
\label{sec:dissociation}

The sublimated molecules are 50\% water by mass \citep{delsemme88}.
The destruction of these water molecules is mostly by
photodissociation:
\begin{eqnarray}
  \mathrm{H_2O + h}\nu & \rightarrow & \mathrm{OH + H + 3.42~eV} \\
              & \rightarrow & \mathrm{H_2 + O(^1}D) + 3.84~\mathrm{eV} \\
              & \rightarrow & \mathrm{H + H + O + 0.7~eV} \\
              & \rightarrow & \mathrm{H_2O^+ + e + 12.4~eV}.
\end{eqnarray}
The first of these reactions is the most likely, with a branching ratio of
86\%.   The reaction rate of this process at 1~AU is given by \citet{combi04}
as $1.04\times 10^{-5}$~s$^{-1}$ for the nonflaring Sun at medium activity. 
The rates given by \citet{combi04} are in broad agreement with those of
previous work by \citet{huebner92}.
If we
assume this rate scales as the inverse square of the distance to the radiation
source, then the photodissociation rate of the comet at the solar surface is 
$\sim 4.8\times 10^{-1}$~s$^{-1}$. This translates to a lifetime of $\sim 2.08$~s
for a water molecule at the solar surface.  At small heliocentric distances,
thermally excited states in the molecule will increase the photodissociation rate
such that the $r^{-2}$ scaling is not entirely accurate. Our estimate of the
lifetime is thus an upper limit (i.e., a lower limit of the rate constant).

The hydroxyl radical resulting from the dissociation of water will also
photodissociate:
\begin{eqnarray}
  \mathrm{OH + h}\nu & \rightarrow & \mathrm{O(^3}P) + \mathrm{H + 1.27~eV} \\
            & \rightarrow & \mathrm{O(^1}D) + \mathrm{H + 7.90~eV} \\
	    & \rightarrow & \mathrm{O(^1}S) + \mathrm{H + 9.80~eV} \\
            & \rightarrow & \mathrm{OH^+ + e + 19.1~eV}.
\end{eqnarray}
To find the total rate for OH dissociating to O and H, we take the sum of the
first three of these reactions. The total rate constant at 1~AU is given by
\citet{combi04} as $2.5\times 10^{-5}$~s$^{-1}$.  
The resulting rate coefficient at the solar surface is then
1.2~s$^{-1}$, corresponding to a lifetime of 0.86~s.
The mean time for a water molecule to dissociate to O and H can then be estimated
as 2.9~s.

In addition to water, various other molecules are sublimated from the surface.
Under typical solar conditions, most of the AIA channels are dominated by
emission from Fe ions so we are interested in any Fe sublimated from the comet
that may contribute to the emission detected by AIA. 
Fe I emission has been observed in the spectra of comet Ikeya-Seki \citep{preston67},
most likely produced by ablation or vaporization of refractory grains.
At the end stages of the asteroidal evaporation, the most important molecule
containing Fe is ferrous oxide (FeO).
The photodissociation cross section for FeO at 252~nm is $1.2\times
10^{-18}$~cm$^2$ \citep{chestakov05}.  The solar flux at 1~AU is $\sim 5\times
10^{12}$~photons~cm$^{-2}$~s$^{-1}$  at 252~nm, so the rate at 1~AU is at least
$\sim 6\times 10^{-6}$~s$^{-1}$.  At the solar surface this is 0.28~s$^{-1}$, or a
lifetime of 3.6~s, i.e., comparable to that of water.
Other molecules, such as MgO, will also rapidly photodissociate, adding another O atom to the cometary
debris.  However, the metallic ion so produced does not emit in the AIA passbands.

\subsection{Ionization}
\label{sec:ionization}

The neutral atoms formed by dissociation are ionized through charge exchange with
coronal protons, through impact with coronal electrons and protons, and by
photoionization. We consider each of these processes in turn.

\subsubsection{Charge exchange}

Electron capture by hydrogen ions due to collisions with neutral oxygen,
\begin{equation}
  \mathrm{H^+ + O(^3}P) \rightarrow \mathrm{H + O^+(^4}S), \label{eqn:cxo1}
\end{equation}
is a near resonant reaction.  Cross sections have been calculated by
\citet{stancil99} for collision energies between 0.1~meV/u and 10~MeV/u. For a
collision energy of 2~keV (corresponding to the energy of H$^+$ in the collision 
with the comet material) the rate
coefficient for the reaction is $5.52\times
10^{-8}$. For a coronal
proton density of $10^8$~cm$^{-3}$, the lifetime of cometary O atoms is thus
0.18~s. The reverse reaction is considerably less likely due to the negligible
abundance of neutral H in the corona.

The resonant charge exchange process between H and protons,
\begin{equation}
  \mathrm{H^+ + H(1}s)  \rightarrow  \mathrm{H(1}s) + \mathrm{H^+} \label{eqn:cxh},
\end{equation}
has been studied by \citet{bates53}. They provide tabulated cross sections down
to an impact energy of 2.5~keV.
The cross section has a weak energy dependence and no other resonances at these energies so 
we can extrapolate below this energy to find a cross section of 
$\sim 6\times 10^{-15}$~cm$^2$ at an energy of 2~keV.  For a
coronal proton density of $10^8$~cm$^{-3}$ and a collision energy of 2~keV, this
translates to a mean lifetime of only 0.027~s before cometary hydrogen atoms are
ionized by charge transfer.  As above, the inverse process is much less efficient
because of the scarcity of neutral H in the corona.

During the initial expansion the density of H and O exceeds $10^8$~cm$^{-3}$.   This means the
time to become ionized is increased until the comet material has expanded to a radius that
includes sufficient protons to ionize all H and O.  Until then, the situation is more
complicated.

\subsubsection{Electron-impact ionization}

Following the initial ionization of the neutral species, subsequent ionization is
dominated by impact with free electrons of the corona.  We estimate the
timescales of these reactions under the assumption that the electron velocity is
large in comparison to that of the cometary ions.  For a Maxwellian distribution
of free electrons with temperature $1.5\times 10^6$~K, the mean electron velocity
is,
\begin{equation}
  v_{\rm e} = \sqrt{\frac{8kT_{\rm e}}{\pi m_{\rm e}}} \approx 7.6\times10^3 \, {\rm km\, s}^{-1},
\end{equation}  
so the free electrons in the corona are over an order of magnitude faster than the
cometary oxygen ions.  For comparison, coronal protons and oxygen ions have mean
velocities of $1.8\times 10^2$ and 44~km~s$^{-1}$, respectively. A proper
analysis of the electron-impact ionization rate would consider an anisotropic
electron velocity distribution in the rest frame of the ion. However, given the
disparity in the velocity of the interacting electrons and ions, it is safe to
proceed under the assumption of an isotropic Maxwellian electron distribution of
$T_{\rm e}=1.5\times 10^6$~K in the rest frame of the target ion.

To trace the ionization level of cometary oxygen as it passes through the corona,
we consider the reactions:
\begin{equation}
  \mathrm{O}^{q+} + \mathrm{e} \rightarrow \mathrm{O}^{(q+1)+} + \mathrm{e + e},
\end{equation}
where $q$ is the charge of the ion.  We use the rate coefficients of
\citet{dere07}, an electron temperature of $1.5\times 10^6$~K, and a density of
$10^8$~cm$^{-3}$ to derive reaction rates for this process. The results are
summarized in Table~\ref{table:ionization}. 

\begin{deluxetable*}{cccccc}
\tablecolumns{6}
\tablewidth{0pt}
\tabletypesize{\small}
\tablecaption{Ionization Rates for Selected Ions (s$^{-1}$)}
\tablehead{
\colhead{Ionization Level} & 
\colhead{CX} & 
\colhead{EII} &
\colhead{Photoionization} & 
\colhead{PII} &
\colhead{Total}}
\startdata
$\mathrm{H} \rightarrow \mathrm{H}^{+}$       & 37.6&  2.9                 & $4.3\times 10^{-3}$ & --                   & 40.5               \\
$\mathrm{O} \rightarrow \mathrm{O}^{+}$       & 5.5 &  8.3                 & $1.5\times 10^{-2}$ & $4.2\times 10^{-1}$  & 14.2                \\
$\mathrm{O}^{+} \rightarrow \mathrm{O}^{2+}$  & --  &  2.8                 & $6.5\times 10^{-3}$ & $3.1\times 10^{-4}$  & 2.8		      \\
$\mathrm{O}^{2+} \rightarrow \mathrm{O}^{3+}$ & --  &  1.2                 & $1.4\times 10^{-3}$ & $4.2\times 10^{-5}$  & 1.2		      \\
$\mathrm{O}^{3+} \rightarrow \mathrm{O}^{4+}$ & --  &  $4.1\times 10^{-1}$ & $5.1\times 10^{-5}$ & $8.0\times 10^{-6}$  & $4.1\times 10^{-1}$ \\
$\mathrm{O}^{4+} \rightarrow \mathrm{O}^{5+}$ & --  &  $1.1\times 10^{-1}$ & --                  & $1.5\times 10^{-6}$  & $1.1\times 10^{-1}$ \\
$\mathrm{O}^{5+} \rightarrow \mathrm{O}^{6+}$ & --  &  $3.4\times 10^{-2}$ & --                  & $3.5\times 10^{-7}$  & $3.4\times 10^{-2}$ \\
$\mathrm{Fe} \rightarrow \mathrm{Fe}^{+}$       & --  &  20.6              & $8.4\times 10^{-2}$                  & 1.7  &  22.3 \\
$\mathrm{Fe}^{+} \rightarrow \mathrm{Fe}^{2+}$  & --  &  6.3		   & --                  & $3.6\times 10^{-3}$  & 6.3		     \\
$\mathrm{Fe}^{2+} \rightarrow \mathrm{Fe}^{3+}$ & --  &  6.6		   & --                  & $5.4\times 10^{-4}$  & 6.6		     \\
$\mathrm{Fe}^{3+} \rightarrow \mathrm{Fe}^{4+}$ & --  &  3.8               & --                  & $8.2\times 10^{-5}$  & 3.8		     \\
$\mathrm{Fe}^{4+} \rightarrow \mathrm{Fe}^{5+}$ & --  &  1.6               & --		         & $1.9\times 10^{-5}$  & 1.6		     \\
$\mathrm{Fe}^{5+} \rightarrow \mathrm{Fe}^{6+}$ & --  &  $8.4\times 10^{-1}$ & --		 & $5.5\times 10^{-6}$  & $8.4\times 10^{-1}$ \\
$\mathrm{Fe}^{6+} \rightarrow \mathrm{Fe}^{7+}$ & --  &  $4.6\times 10^{-1}$ & --                & $1.9\times 10^{-6}$  & $4.6\times 10^{-1}$ \\
$\mathrm{Fe}^{7+} \rightarrow \mathrm{Fe}^{8+}$ & --  &  $2.0\times 10^{-1}$ & --                & $6.8\times 10^{-7}$  & $2.0\times 10^{-1}$ \\
$\mathrm{Fe}^{8+} \rightarrow \mathrm{Fe}^{9+}$ & --  &  $5.4\times 10^{-2}$ & --		 & $3.0\times 10^{-7}$  & $5.4\times 10^{-2}$ \\
$\mathrm{Fe}^{9+} \rightarrow \mathrm{Fe}^{10+}$ & -- &  $3.0\times 10^{-2}$ & --		 & $1.7\times 10^{-7}$  & $3.0\times 10^{-2}$ \\
$\mathrm{Fe}^{10+} \rightarrow \mathrm{Fe}^{11+}$ & --&  $1.5\times 10^{-2}$ & --		 & $9.5\times 10^{-8}$  & $1.5\times 10^{-2}$ \\
$\mathrm{Fe}^{11+} \rightarrow \mathrm{Fe}^{12+}$ & --&  $7.5\times 10^{-3}$ & --                & $4.8\times 10^{-8}$  & $7.5\times 10^{-3}$ \\
$\mathrm{Fe}^{12+} \rightarrow \mathrm{Fe}^{13+}$ & --&  $4.0\times 10^{-3}$ & --                & $2.6\times 10^{-8}$  & $4.0\times 10^{-3}$ 
\enddata											   
\tablecomments{Ionization rates of various processes at the solar surface.		   
Charge exchange (CX), electron-impact (EII) and proton-impact ionization (PII) rates are for	   
an electron and proton density of $10^8$~cm$^{-3}$.
Photoionization was not considered for states O$^{4+}$ and higher, or Fe$^{+}$ and higher.}						   
\label{table:ionization}									   
\end{deluxetable*}	

From these results we can estimate which ionization stages of O will be
sufficiently long-lived to contribute to EUV emission. If the plasma were allowed
to relax to ionization equilibrium then the highest ionization stage reached
under these conditions would be O$^{6+}$ 
\citep{bryans09}.  We thus expect to see emission from ionization stages up to
and including $q=6$.  Given the ionization rates 
 of the cometary material, this ionization state will be reached only after sufficient 
 time has elapsed.

\citet{dere07} also provide rate coefficients for the electron-impact ionization
of Fe. We have given the rates for the first 12 ionization stages at an electron
temperature of $1.5\times 10^6$~K and a density of $10^8$~cm$^{-3}$ in
Table~\ref{table:ionization}. Analyzing these timescales shows that the highest
ionization stage reached is  Fe$^{9+}$ for $v_{\rm out}=17$~km~s$^{-1}$.

\subsubsection{Photoionization}

The photoionization rate of the dissociated atoms and ions is dependent on the
solar EUV flux.  The comet is exposed to a wide range of radiances as it travels
close to the solar surface.  For this reason, we choose to use the average EUV
spectral irradiances at 1~AU from the Thermosphere Ionosphere Mesosphere
Energetics Dynamics (TIMED) spacecraft's Solar EUV Experiment
\citep[SEE;][]{see}.  Values between 2002 and 2011.5 were averaged to produce
the irradiances used here. We use the cross sections from \citet{verner95} and
\citet{verner96} to calculate the photoionization rates at 1~AU and have
calculated the equivalent rate at the solar surface by scaling by $r^2$. The
values are given in Table~\ref{table:ionization}. In comparison to resonant
charge exchange and electron-impact ionization, photoionization has a negligible
impact on the ionization rate. 
A flare could have larger
photoionization consequences, but none were observed during the perihelion
passage of either of the comets discussed here.
For this reason we have only listed the
photoionization rate for the low charge  states of O and Fe in
Table~\ref{table:ionization}.

Photoionization of neutral hydrogen also plays a role in the production of
protons. According to \citet{keller76}, the calculated lifetime of cometary H
atoms due to photoionization from the solar radiation field at a distance of 1~AU
is $1.4\times 10^7$~s.   This value is comparable to that from the TIMED/SEE
irradiances, where the average photoionization lifetime over 2002--2011.5
was calculated as $1.1\times 10^{7}$~s at 1~AU. The resulting
lifetime against photoionization at the solar surface is $2.4\times 10^2$~s.

\subsubsection{Ionization by protons}

The relative velocity of the cometary ions to the protons in the corona is $\sim
600$~km~s$^{-1}$.  To calculate the rate of ionization due to impact with
energetic protons we use the classical approximation of \citet{gryzinski65}.
Results are shown in Table~\ref{table:ionization} for O and Fe.  They are
negligible compared to other ionization processes.

Cometary protons, resulting from sublimated water molecules that have been
dissociated and subsequently ionized, will initially be traveling at the velocity
of the comet relative to the corona.  This velocity of 600~km~s$^{-1}$
corresponds to a proton beam of 2~keV.  Can these fast protons cause ionization
and excitation of the coronal plasma they are passing through?  This is the same
reaction as described above where fast cometary ions collide with slower coronal
protons.  Using the same approach, we can show that the  proton impact does not
have a significant effect on the ionization balance of the plasma.

For it to be important, the proton-impact rate has to be comparable to the
electron-impact ionization rate and the recombination rate.  Taking ionization of
O as an example, the rate coefficient for proton impact is $\sim 5$ orders of
magnitude smaller than that for electron impact for the ions present in the
corona (see Table~\ref{table:ionization}). For the conditions outlined in
Section~\ref{sec:model} and the dissociation time for water given in
Section~\ref{sec:dissociation}, the highest proton density we can expect in the
comet is $\sim 10^{11}$~cm$^{-3}$, only around 3 orders of magnitude greater than
the coronal density.  The ionization rate by proton impact is thus 2 orders of
magnitude smaller than that by electron impact.

\section{Results}
\label{sec:results}

\begin{figure}
  \centering
  \includegraphics[width=0.5\textwidth]{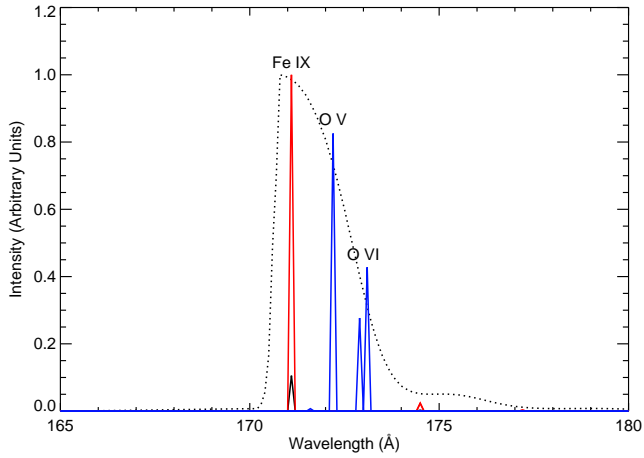}
  \caption{Emission from cometary ions in the bandpass of AIA 171~\AA\ compared to QS. 
Fe emission from the comet is shown in red and O emission from the comet in blue.
Black lines indicate a typical spectrum due to the QS. 
The dotted line is the effective area of the AIA passband. 
The colored lines are the spectrum from cometary O and Fe ions with
outgassing velocity of 17~km~s$^{-1}$.
The strongest emission lines have been labeled.}
  \label{fig:171}
\end{figure}

\begin{figure}
  \centering
  \includegraphics[width=0.5\textwidth]{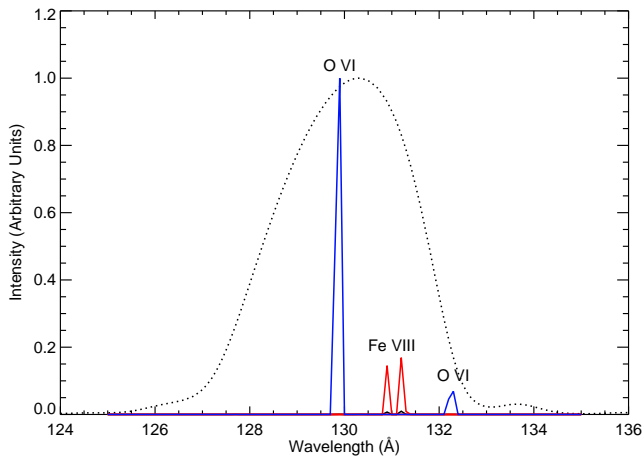}
  \caption{As Fig.~\protect\ref{fig:171} but for 131~\AA.}
  \label{fig:131}
\end{figure}

\begin{figure}
  \centering
  \includegraphics[width=0.5\textwidth]{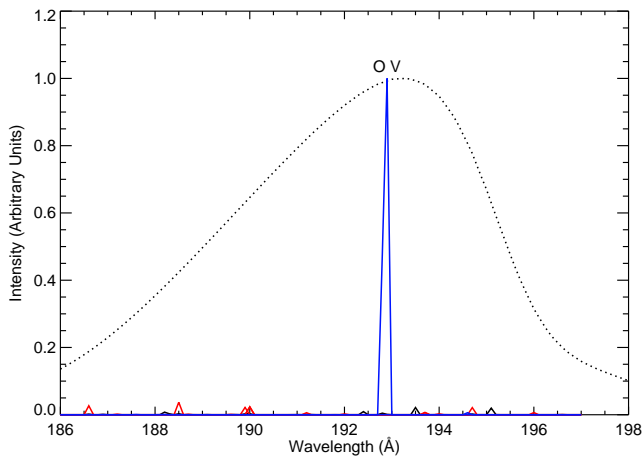}
  \caption{As Fig.~\protect\ref{fig:171} but for 193~\AA.}
  \label{fig:193}
\end{figure}

\begin{figure}
  \centering
  \includegraphics[width=0.5\textwidth]{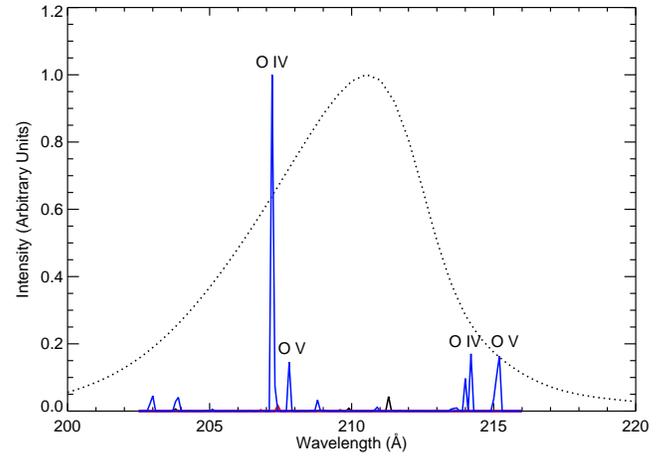}
  \caption{As Fig.~\protect\ref{fig:171} but for 211~\AA.}
  \label{fig:211}
\end{figure}

\begin{figure}
  \centering
  \includegraphics[width=0.5\textwidth]{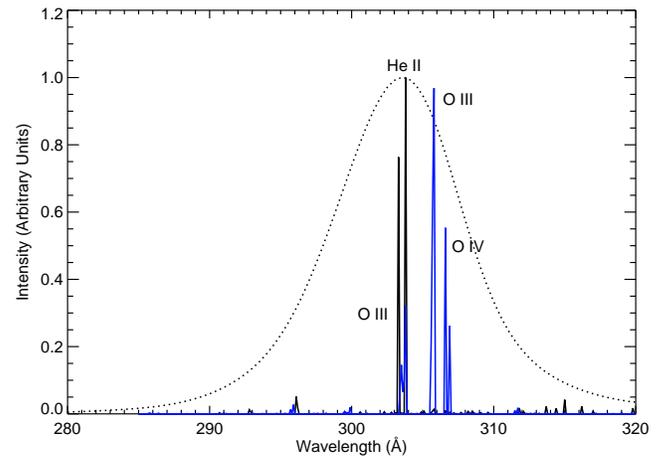}
  \caption{As Fig.~\protect\ref{fig:171} but for 304~\AA.}
  \label{fig:304}
\end{figure}

\begin{figure}
  \centering
  \includegraphics[width=0.5\textwidth]{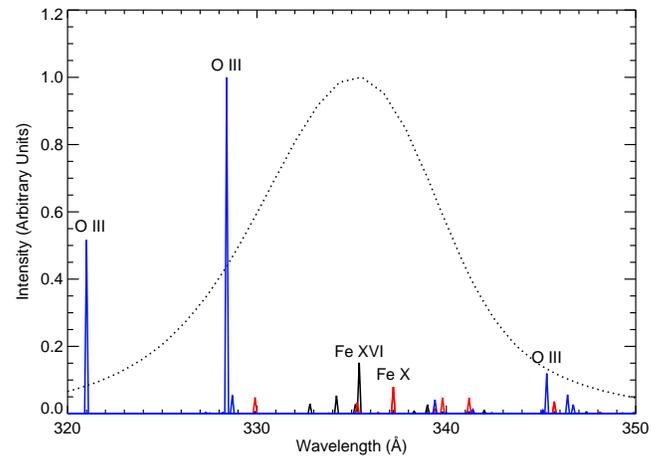}
  \caption{As Fig.~\protect\ref{fig:171} but for 335~\AA.}
  \label{fig:335}
\end{figure}

\begin{figure}
  \centering
  \includegraphics[width=0.5\textwidth]{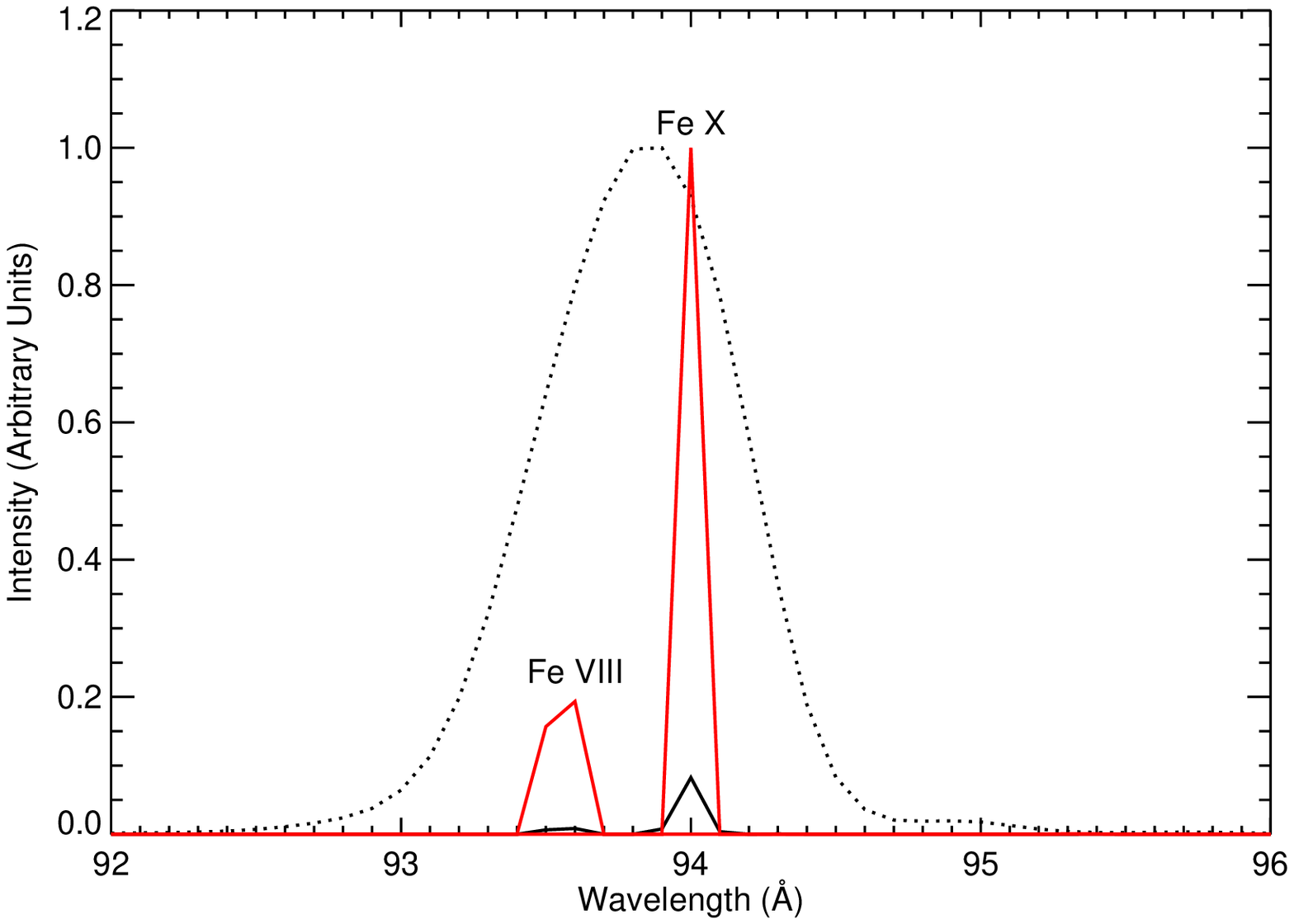}
  \caption{As Fig.~\protect\ref{fig:171} but for 94~\AA.}
  \label{fig:94}
\end{figure}

Using the physical conditions outlined in Section~\ref{sec:model} and the ionization rates
of Section~\ref{sec:ionization}, we have calculated the line emission from all elements
listed in Table~\ref{table:abund} in each AIA bandpass using CHIANTI \citep{chianti1,
chianti6}.  Collisional redistribution among the excited states of the ion is several orders
of magnitude faster than ionization, so we have assumed detailed balance applies to the
distribution of those energy levels.  The results are shown in
Figures~\ref{fig:171}--\ref{fig:94} for an outflow time of 60~s.  The intensities have been
folded with the effective areas of the respective AIA filters \citep[taken from SolarSoft; see also][] {boerner12} and compared
to the average QS emission.  All intensities are integrated over the emitting area
described in Section~\ref{sec:model}  and outlined in Fig.~\ref{fig:cartoon} and compared
with QS intensities over the same area. Of the elements considered (H, O, C, N, Mg,
Si, S, and Fe), only the ions of O and Fe contribute any  significant emissions in the
wavelength ranges of the AIA instrument.

We note here that the He~{\sc ii} lines that dominate the QS emission in the 304~\AA\ channel 
are not properly modeled by CHIANTI, which underestimates the observed intensities by a
significant factor \citep[see, e.g.,][]{andretta03}.  For this reason, we have 
multiplied the background QS emission from these lines by a factor of 20.0 to bring them in
line with observation \citep{warren05}.

\begin{figure}
  \centering
  \includegraphics[width=0.5\textwidth]{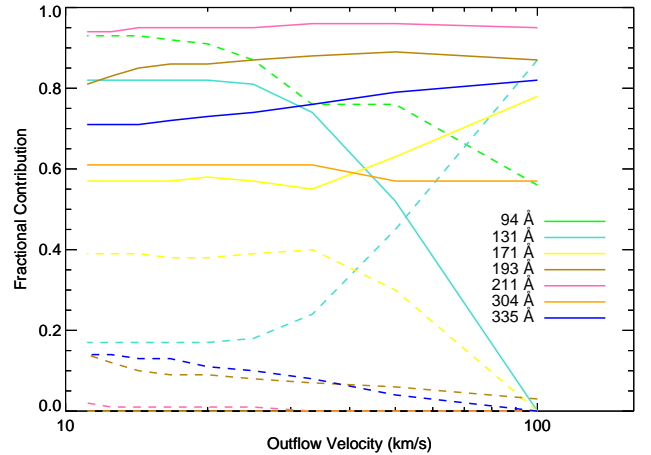}
  \caption{The intensity of O and Fe emission as a function of outflow velocity
   as detected by each AIA bandpass is given as a fraction of the total emission 
   from the comet plus the average QS emission.  The results are given for the
   conditions given in Section~\ref{sec:model}. Solid lines show the fractional contribution
   from cometary O ions and dashed lines from cometary Fe ions.}
  \label{fig:frac}
\end{figure}

For each AIA bandpass we have calculated the fractional increase in radiance it would
detect due to the cometary emission mechanism. We have
integrated the emission over the wavelength range of each bandpass and compared
the total intensity of the QS plus comet emission to QS emission
alone.  We have calculated the dependence of this fractional increase on the outflow
velocity and show the results in Fig.~\ref{fig:frac}.  The time for the cometary material to
flow from the nucleus to the edge of the emitting region is allowed the range 10--90~s (an
outflow velocity of 11--100~km~s$^{-1}$).
The solid lines show the contribution from O and the dashed lines show the contribution from
Fe.
The outflow velocity has an impact on the relative contribution from O and Fe ions to the emission spectra, most notably in the 131 and 171~\AA\ passbands, but does not alter the total fraction of emission from cometary ions relative to the background to a significant degree.

Our model predicts emission in the 171~\AA\ bandpass from O~{\sc v}, O~{\sc vi}, Fe~{\sc ix}, and a smaller
contribution from Fe~{\sc x}.  Under normal QS conditions this bandpass is dominated by
Fe~{\sc ix} emission.  The only other bandpass with significant emission from both O and Fe due to
the cometary material is the
131~\AA\ channel.  Here, contributions to the emission come from O~{\sc vi} and Fe~{\sc viii}.  These
same Fe~{\sc viii} lines are those normally strongest in this channel under QS  conditions. 
It is significant that the 171 and 131~\AA\ channels are the only two that detect strong O~{\sc vi}
emission lines.  These are the highest ionization stages of O detectable in any of the AIA
passbands, taking longer to form than the lower stages, and thus explaining the time delay in
the observations from these channels (see Fig.~\ref{fig:lovejoy ingress}).  We expand on this
argument in Section~\ref{sec:discussion}.

In the 193, 211, 304, and 335~\AA\  passbands, O emission dominates over Fe. A single O~{\sc v}
line dominates the emission spectrum in the 193~\AA\ passband. The 211~\AA\ passband has
contributions from a number of O~{\sc iv} and O~{\sc v} lines. O~{\sc iii} and O~{\sc iv} are the strongest emitters
in the 304~\AA\ passband, and O~{\sc iii} dominates the emission in the 335~\AA\ passband.  We
note a progression of emission from O~{\sc v}, through O~{\sc iv}, to O~{\sc iii} on advancing through the
193, 211, 304, and 335~\AA\  passbands. We return to this point in
Section~\ref{sec:discussion} when discussing a possible time lag in the observations from
these channels.

The 94~\AA\ passband has no O emission lines in its wavelength range. Here, emission from
the comet material is from Fe~{\sc viii} and Fe~{\sc x} and acts to increase the same emission lines
that are strongest in the quiescent QS.

\section{Discussion and Conclusions}
\label{sec:discussion}

We have shown that emission from the O and Fe atoms deposited in the corona 
by the passage of a comet is a viable mechanism to
explain the increased radiance detected by AIA.  The rapid injection of
atoms to the corona followed by their ionization through successive ionization
stages produces emission lines not normally present at coronal temperatures.
The exact structure of the resulting spectra and their magnitude relative to the
background corona are dependent on both the coronal and cometary parameters. 
Comparing our calculations with the AIA observations may provide a means of
determining these parameters more accurately than has been possible heretofore.

The difference in the emission detected across the AIA channels for C/2011 W3 (Lovejoy) can
be explained by our model if one considers the O emission.  There is some observational
evidence of the 304 and 335~\AA\ channels showing emission only close to the comet nucleus
but that does not persist.  By contrast, the 131 and 171~\AA\ channels show more spatially
extended emission that persists for up to 20 minutes.  The 304~\AA\ channel has contributions
from O~{\sc iii} and O~{\sc iv} while 335~\AA\ has strong contributions from O~{\sc iv}.  According to our
model, O will emit in these ionization stages for several seconds before being further
ionized.  Both 131 and 171~\AA\ channels have significant contributions from O~{\sc vi}.  This ion
has a higher formation temperature, close to the ambient coronal temperature, and so remains
present in the plasma for longer than the lower ionization stages.
The observed fading must come from the diminution in the number density of oxygen nuclei
in the field-of-view.

By a similar argument, the model presented here predicts that there should be a time lag in
the emission from the AIA channels.  This should follow the ionization stages of O that
contribute to the emission in each wavelength channel, i.e., emission should first be
detected in 335, followed by 304, then 211 and 193, and finally in 131 and 171~\AA. 
Lightcurves of C/2011 W3 (Lovejoy) are being measured and the time lag between the
channels is expected to be only a few seconds (see Table~\ref{table:ionization}) so it
remains to be seen whether observations 
spaced by 12 seconds can resolve any such time delay.

Estimates of several parameters influence the predicted emission to varying
degrees.  The assumed density of the corona through which the comet passes is
particularly important.  A larger density than that assumed here
($10^8$~cm$^{-3}$) would result in faster ionization and hence increased emission
from higher ionization stages of O and Fe.  The density assumed for the plasma responsible for
the QS emission
does not affect the cometary emission but will change the spectrum of the
background corona that interferes with the detection of the comet.
Finally, the density of the outgassing molecules will impact
the relative intensity of the cometary and QS spectra.  Within the range
of reasonably expected conditions (around 1--2$\times 10^6$~K), the temperature assumed for the corona does
not significantly alter the calculated spectra.

Our model assumes the cometary species move radially away from the nucleus of the
comet after sublimating.  This assumption neglects the influence of the
magnetic field on the ions.  The primary effect of the gyrorotation will be an 
increase in the effective
pathlength of the ions as they spiral along the field lines.  A proper analysis
would require the direction and magnitude of the magnetic field, and is beyond
the scope of this work, but is worth further study.

Intensity profiles from AIA channels also show indications of absorption by the
denser inner regions of the comet's coma when viewed against the disk of the Sun
 \citep{schrijver12}.  The model
presented here may be able to account for this.  Near the comet surface, ions
have yet to be ionized to the stages responsible for the emission seen later in
the AIA observations.  It is possible that the lower ionization stages cause
absorption in this region and the relative brightness may track the O~{\sc i} photoionization curve. We intend to explore this possibility in future work.

\acknowledgements

Research was supported by the Solar Dynamics Observatory.

We acknowledge many helpful comments and suggestions from K. Battams, J.~C. Brown, H.~S. Hudson, W. Liu, P. Saint-Hilaire, and C.~J. Schrijver.

\bibliographystyle{apj}
\bibliography{refs}

\begin{thebibliography}{39}
\expandafter\ifx\csname natexlab\endcsname\relax\def\natexlab#1{#1}\fi

\bibitem[{{Andretta} {et~al.}(2003){Andretta}, {Del Zanna}, \&
  {Jordan}}]{andretta03}
{Andretta}, V., {Del Zanna}, G., \& {Jordan}, S.~D. 2003, \aap, 400, 737

\bibitem[{{Bates} \& {Dalgarno}(1953)}]{bates53}
{Bates}, D.~R., \& {Dalgarno}, A. 1953, Proceedings of the Physical Society A,
  66, 972

\bibitem[{{Biesecker} {et~al.}(2002){Biesecker}, {Lamy}, {St.~Cyr}, {Llebaria},
  \& {Howard}}]{biesecker02}
{Biesecker}, D.~A., {Lamy}, P., {St.~Cyr}, O.~C., {Llebaria}, A., \& {Howard},
  R.~A. 2002, \icarus, 157, 323

\bibitem[{{Boerner} {et~al.}(2012){Boerner}, {Edwards}, {Lemen}, {Rausch},
  {Schrijver}, {Shine}, {Shing}, {Stern}, {Tarbell}, {Title}, {Wolfson},
  {Soufli}, {Spiller}, {Gullikson}, {McKenzie}, {Windt}, {Golub}, {Podgorski},
  {Testa}, \& {Weber}}]{boerner12}
{Boerner}, P., {et~al.} 2012, \solphys, 275, 41

\bibitem[{{Brown} {et~al.}(2011){Brown}, {Potts}, {Porter}, \& {Le
  Chat}}]{brown11}
{Brown}, J.~C., {Potts}, H.~E., {Porter}, L.~J., \& {Le Chat}, G. 2011, \aap,
  535, A71

\bibitem[{{Brueckner} {et~al.}(1995){Brueckner}, {Howard}, {Koomen},
  {Korendyke}, {Michels}, {Moses}, {Socker}, {Dere}, {Lamy}, {Llebaria},
  {Bout}, {Schwenn}, {Simnett}, {Bedford}, \& {Eyles}}]{lasco}
{Brueckner}, G.~E., {et~al.} 1995, \solphys, 162, 357

\bibitem[{{Bryans} {et~al.}(2009){Bryans}, {Landi}, \& {Savin}}]{bryans09}
{Bryans}, P., {Landi}, E., \& {Savin}, D.~W. 2009, \apj, 691, 1540

\bibitem[{{Chestakov} {et~al.}(2005){Chestakov}, {Parker}, \&
  {Baklanov}}]{chestakov05}
{Chestakov}, D.~A., {Parker}, D.~H., \& {Baklanov}, A.~V. 2005, \jcp, 122,
  084302

\bibitem[{{Combi} {et~al.}(2004){Combi}, {Harris}, \& {Smyth}}]{combi04}
{Combi}, M.~R., {Harris}, W.~M., \& {Smyth}, W.~H. 2004, {Comets II. Gas
  dynamics and kinetics in the cometary coma: theory and observations}, ed.
  {Festou, M.~C., Keller, H.~U., \& Weaver, H.~A.}, 523--552

\bibitem[{{Cravens}(1997)}]{cravens97}
{Cravens}, T.~E. 1997, \grl, 24, 105

\bibitem[{{Delsemme}(1988)}]{delsemme88}
{Delsemme}, A.~H. 1988, Royal Society of London Philosophical Transactions
  Series A, 325, 509

\bibitem[{Dennerl {et~al.}(1997)Dennerl, Englhauser, \& Trumper}]{dennerl97}
Dennerl, K., Englhauser, J., \& Trumper, J. 1997, Science, 277, 1625

\bibitem[{{Dere}(2007)}]{dere07}
{Dere}, K.~P. 2007, \aap, 466, 771

\bibitem[{{Dere} {et~al.}(1997){Dere}, {Landi}, {Mason}, {Monsignori Fossi}, \&
  {Young}}]{chianti1}
{Dere}, K.~P., {Landi}, E., {Mason}, H.~E., {Monsignori Fossi}, B.~C., \&
  {Young}, P.~R. 1997, \aaps, 125, 149

\bibitem[{{Dere} {et~al.}(2009){Dere}, {Landi}, {Young}, {Del Zanna},
  {Landini}, \& {Mason}}]{chianti6}
{Dere}, K.~P., {Landi}, E., {Young}, P.~R., {Del Zanna}, G., {Landini}, M., \&
  {Mason}, H.~E. 2009, \aap, 498, 915

\bibitem[{{Domingo} {et~al.}(1995){Domingo}, {Fleck}, \& {Poland}}]{soho}
{Domingo}, V., {Fleck}, B., \& {Poland}, A.~I. 1995, \solphys, 162, 1

\bibitem[{{Green}(2011)}]{green11}
{Green}, D.~W.~E. 2011, Central Bureau Electronic Telegrams, 2930

\bibitem[{{Gryzi{\'n}ski}(1965)}]{gryzinski65}
{Gryzi{\'n}ski}, M. 1965, Physical Review, 138, 336

\bibitem[{{Gundlach} {et~al.}(2012){Gundlach}, {Blum}, {Skorov}, \&
  {Keller}}]{gundlach12}
{Gundlach}, B., {Blum}, J., {Skorov}, Y.~V., \& {Keller}, H.~U. 2012, ArXiv
  e-prints

\bibitem[{{Huebner} {et~al.}(1992){Huebner}, {Keady}, \& {Lyon}}]{huebner92}
{Huebner}, W.~F., {Keady}, J.~J., \& {Lyon}, S.~P. 1992, \apss, 195, 1

\bibitem[{{Keller}(1976)}]{keller76}
{Keller}, H.~U. 1976, \ssr, 18, 641

\bibitem[{{Knight} {et~al.}(2010){Knight}, {A'Hearn}, {Biesecker}, {Faury},
  {Hamilton}, {Lamy}, \& {Llebaria}}]{knight10}
{Knight}, M.~M., {A'Hearn}, M.~F., {Biesecker}, D.~A., {Faury}, G., {Hamilton},
  D.~P., {Lamy}, P., \& {Llebaria}, A. 2010, \aj, 139, 926

\bibitem[{{Krasnopolsky}(1997)}]{krasnopolsky97}
{Krasnopolsky}, V. 1997, \icarus, 128, 368

\bibitem[{{Lemen} {et~al.}(2012){Lemen}, {Title}, {Akin}, {Boerner}, {Chou},
  {Drake}, {Duncan}, {Edwards}, {Friedlaender}, {Heyman}, {Hurlburt}, {Katz},
  {Kushner}, {Levay}, {Lindgren}, {Mathur}, {McFeaters}, {Mitchell}, {Rehse},
  {Schrijver}, {Springer}, {Stern}, {Tarbell}, {Wuelser}, {Wolfson}, {Yanari},
  {Bookbinder}, {Cheimets}, {Caldwell}, {Deluca}, {Gates}, {Golub}, {Park},
  {Podgorski}, {Bush}, {Scherrer}, {Gummin}, {Smith}, {Auker}, {Jerram},
  {Pool}, {Soufli}, {Windt}, {Beardsley}, {Clapp}, {Lang}, \& {Waltham}}]{aia}
{Lemen}, J.~R., {et~al.} 2012, \solphys, 275, 17

\bibitem[{Lisse {et~al.}(1996)Lisse, Dennerl, Englhauser, Harden, Marshall,
  Mumma, Petre, Pye, Ricketts, Schmitt, Trumper, \& West}]{lisse96}
Lisse, C., {et~al.} 1996, Science, 274, 205

\bibitem[{{Marsden}(2005)}]{marsden05}
{Marsden}, B.~G. 2005, \araa, 43, 75

\bibitem[{{Pesnell} {et~al.}(2012){Pesnell}, {Thompson}, \& {Chamberlin}}]{sdo}
{Pesnell}, W.~D., {Thompson}, B.~J., \& {Chamberlin}, P.~C. 2012, \solphys,
  275, 3

\bibitem[{{Preston}(1967)}]{preston67}
{Preston}, G.~W. 1967, \apj, 147, 718

\bibitem[{Schrijver {et~al.}(2012)Schrijver, Brown, Battams, Saint-Hilaire,
  Liu, Hudson, \& Pesnell}]{schrijver12}
Schrijver, C.~J., Brown, J.~C., Battams, K., Saint-Hilaire, P., Liu, W.,
  Hudson, H., \& Pesnell, W.~D. 2012, Science, 335, 324

\bibitem[{{Sekanina}(1984)}]{sekanina84}
{Sekanina}, Z. 1984, \icarus, 58, 81

\bibitem[{{Sekanina}(2003)}]{sekanina03}
---. 2003, \apj, 597, 1237

\bibitem[{{Sekanina} \& {Chodas}(2012)}]{sekanina12}
{Sekanina}, Z., \& {Chodas}, P.~W. 2012, ArXiv e-prints

\bibitem[{{Stancil} {et~al.}(1999){Stancil}, {Schultz}, {Kimura}, {Gu},
  {Hirsch}, \& {Buenker}}]{stancil99}
{Stancil}, P.~C., {Schultz}, D.~R., {Kimura}, M., {Gu}, J.-P., {Hirsch}, G., \&
  {Buenker}, R.~J. 1999, \aaps, 140, 225

\bibitem[{{Vernazza} \& {Reeves}(1978)}]{vernazza78}
{Vernazza}, J.~E., \& {Reeves}, E.~M. 1978, \apjs, 37, 485

\bibitem[{{Verner} {et~al.}(1996){Verner}, {Ferland}, {Korista}, \&
  {Yakovlev}}]{verner96}
{Verner}, D.~A., {Ferland}, G.~J., {Korista}, K.~T., \& {Yakovlev}, D.~G. 1996,
  \apj, 465, 487

\bibitem[{{Verner} \& {Yakovlev}(1995)}]{verner95}
{Verner}, D.~A., \& {Yakovlev}, D.~G. 1995, \aaps, 109, 125

\bibitem[{{Warren}(2005)}]{warren05}
{Warren}, H.~P. 2005, \apjs, 157, 147

\bibitem[{{Weissman}(1983)}]{weissman83}
{Weissman}, P.~R. 1983, \icarus, 55, 448

\bibitem[{{Woods} {et~al.}(2005){Woods}, {Eparvier}, {Bailey}, {Chamberlin},
  {Lean}, {Rottman}, {Solomon}, {Tobiska}, \& {Woodraska}}]{see}
{Woods}, T.~N., {et~al.} 2005, Journal of Geophysical Research (Space Physics),
  110, 1312

\end{thebibliography}

\end{document}